# Hacking Encrypted Wireless Power: Cyber-Security of Dynamic Charging

Hui Wang, Nima Tashakor, Wei Jiang, Wei Liu, C. Q. Jiang, and Stefan M. Goetz

*Abstract*—Recently, energy encryption for wireless power transfer (WPT) has been developed for energy safety, which is important in public places to suppress unauthorized energy extraction. Most techniques vary the frequency so that unauthorized receivers cannot extract energy because of non-resonance. However, this strategy is unreliable. To stimulate the progress of energy encryption technology and point out security holes, this paper proposes a decryption method for the fundamental principle of encrypted frequency-varying WPT. The paper uses an auxiliary coil to detect the frequency and a switched-capacitor array to adaptively compensate the receiver for a wide frequency range. The switched-capacitor array contains two capacitors and one semiconductor switch. One capacitor compensates the receiver all the time while the other's active time during one WPT cycle is regulated by the switch. Thus, the proposed hacking receiver controls the equivalent capacitance of the compensation and steals WPT energy. Finally, a detailed simulation model and experimental results prove the effectiveness of the attack on frequency-hopping energy encryption. Although any nonnegligible energy extracted would be problematic, we achieved to steal 78% to 84% of the energy an authorized receiver could get. When the frequency changes, the interceptor is coarsely tuned very quickly, which can hack fast frequency-varying encrypted system.

## I. INTRODUCTION

WIRELESS power transfer (WPT) is a widely known solution in contactless charging [1]. The main practical advantages compared to traditional wired charging are high flexibility through electromagnetic coupling, on-road move-and-charge ability [2] as well as high safety due to the avoidance of any connector and bare contacts [3]. Such features make WPT very popular for charging smartphones [4], electric motors [5], electric vehicles [6], medical devices [7], and implants [8].

However, despite numerous advantages, energy safety is still a major concern. For a public charging service, unauthorized users also can harvest energy in this electromagnetic field [9].

To solve this problem, various energy encryption methods have been proposed. Static wireless charging with magnetic field editing is desired [10], i.e., the transmitter knows the authorized user's position and selectively charges that area accordingly [11]. Thus, unauthorized users should not have access to the magnetic fields and cannot steal energy, as shown in Fig. 1(a). However, in many applications, it is complicated to shield off the field entirely and make it inaccessible to any form of interceptor; this applies particularly if the receiver should have the freedom to move [12]. For roadway charging, for instance, multiple authorized users drive fast on the road [13]; so all transmitter coils should be activated, and unauthorized users are unavoidably involved.

Therefore, in some public places, power suppliers prefer frequency-varying strategies [14], as shown in Fig. 1(b). In principle, only authorized receivers know the WPT frequency (sequence), which may be the key itself as in digital cyphers, most obviously stream cyphers, or exchanged on a separate secure digital communication channel. Thus, only authorized receivers should be able to tune their resonators through a capacitor to compensate the receiver, while unauthorized users cannot harvest energy because of the impedance mismatch of the receiving circuit [15]. A switched-capacitor array is the most popular compensation for frequency-varying encryption [16], as the transceiver can jump between multiple fixed resonance frequencies [17]. In principle, the number of resonant frequency points is equal to the number of parallel capacitors [18]. Recently developed topologies for capacitor compensation [19], such as higher-order compensation [20] or capacitor matrices [21], can offer more resonant frequency points. Moreover, to complicate energy interception, Qi et al. presented a stepless frequency compensation method, which can control the frequency from 90 kHz to 150 kHz with a variable capacitor [22].

However, we will demonstrate that encryption through frequency-hopping or -varying is not reliable and can be hacked easily. To avoid someone abusing this energy encryption method and to stimulate more researchers to pay attention to energy encryption, this paper demonstrates an energy decryption attack on frequency-varying WPT systems. The ingredients are an auxiliary coil to detect the WPT frequency in time and a continuous switched-capacitor array to compensate the receiver for a wide frequency range.

This paper is organized as follows: Section II will first analyze the system configuration. Next, Section III will present the frequency detection and the stepless frequency compensation. Section IV follows with the system design procedure. A series of computer simulations and experiments verify the approach in Section V and Section VI, respectively. Finally, Section VII summarizes the paper.

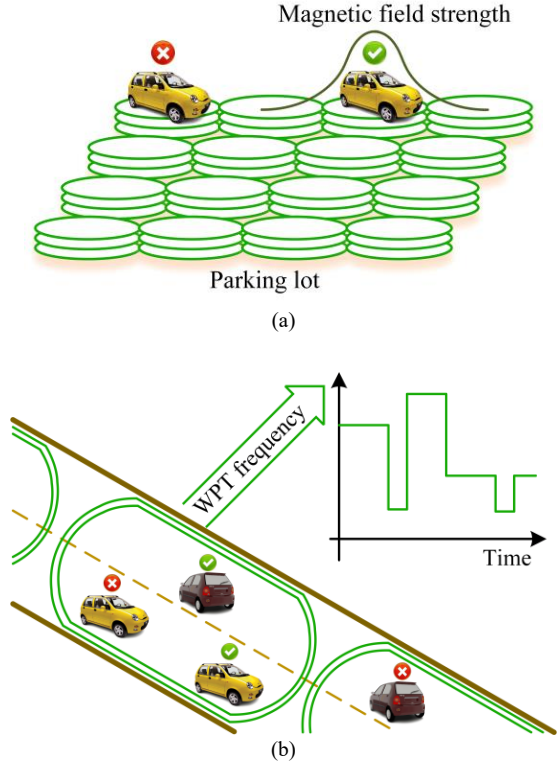

Fig. 1. Current energy protection methods for WPT. (a) Magnetic field editing for static charging. (b) Frequency-variation for dynamic charging.

## II. SYSTEM CONFIGURATION

As shown in Fig. 2, both authorized and unauthorized receivers have access to the electromagnetic field created by the transmitter and attempt to harvest energy, where $L_T$ is the inductance of the transmitter, while $L_R$ and $L_A$ denote the inductances of the hacking receiver and auxiliary coil, respectively; $I_T$ and $I_R$ are currents of transmitter and hacking receiver, respectively; $M_R$ denotes the mutual inductance between transmitter and receiver, and $M_A$ denotes the mutual inductance between transmitter and auxiliary coil, while $M_{RA}$ is the mutual inductance between hacking receiver and auxiliary coil; $C_{R1}$ and $C_{R2}$ are the capacitors to compensate $L_R$ for a wide frequency range, and $S_R$ is the switch to control $C_{R2}$; $V_{CR1}$, $V_{CR2}$, $V_{SR}$, $I_{R1}$, and $I_{R2}$ are the voltages and currents of corresponding capacitors and switch, respectively; $V_{LR}$ and $V_{LA}$ are the voltages of $L_R$ and $L_A$, respectively; $V_{RL}$ is the load voltage.

It should be mentioned that, although the compensation settings of the transmitter and authorized receivers are unknown, the transmitter frequency $f_T$ should be able to vary throughout a wide range [16].

The interceptor coil $L_R$ serves as the receiver to steal wireless power, while the auxiliary coil $L_A$ is a small open-loop sensor coil to detect the phase and frequency of the transmitter current $I_T$.

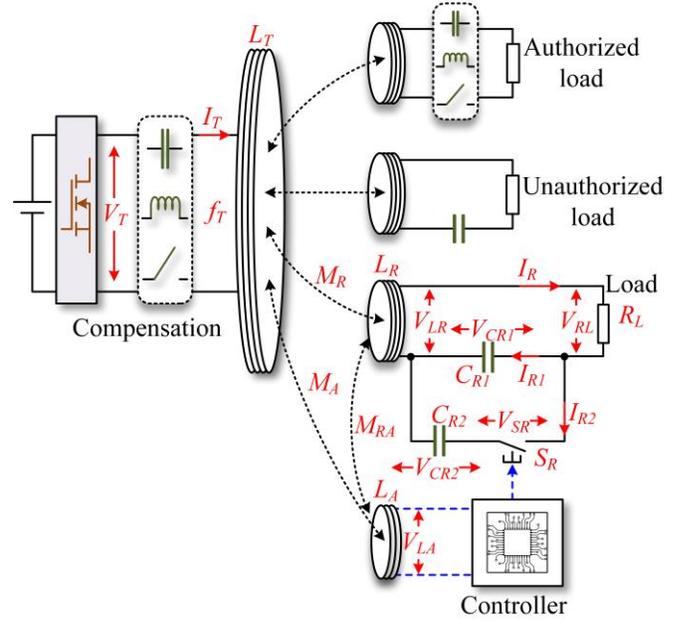

Fig. 2. Common wireless charging system and the proposed hacking receiver.

Besides, it should be mentioned that the compensation network only contains one switch and two capacitors, which is sufficient to compensate the receiver $L_R$ for a wide frequency range. The key is controlling the turn-on time of the switch and adjusting the duty cycle of the controlled capacitor during one cycle. Therefore, the equivalent capacitance of the compensation network can be continuously controlled over a wide range. Also, the compensation network retains the merit of being simple and robust when compared with high-order or capacitor-matrix compensation networks.

## III. SYSTEM DECRYPTION OPERATION STRATEGY

### A. Frequency and Phase Detection

For the proposed energy decryption method, the detection of WPT frequency $f_T$ and phase is the first step in wireless power decryption.

Based on the most basic electromagnetic induction principle of WPT, voltages $V_{LR}$ and $V_{LA}$ can be expressed as [23]

$$\begin{cases} V_{LR} = 2\pi f_T M_R I_T - 2\pi f_R L_R I_R \\ V_{LA} = 2\pi f_T M_A I_T + 2\pi f_R M_{RA} I_R \end{cases}. \quad (1)$$

Obviously, both $V_{LR}$ and $V_{LA}$ contain the frequency and phase information of $I_T$. However, $V_{LR}$ is much easier to be affected by the receiver's current $I_R$, as $L_R$ is much larger than $M_{RA}$. When the receiving circuit of the variable-capacitor compensation is not resonant, $I_R$ is not sinusoidal and strongly distorted. Thus, $V_{LR}$ is disturbed, which would deteriorate any estimation of the phase.

The simple auxiliary coil as a sensor is practically a Kelvin-connected field detection and allows rapid undistorted estimation of field properties [24]. The frequency $f_T$ is narrowed down already after the first zero crossings through simple counting. Also, the voltage upward zero crossings can be detected and treated as zero phase, which will be used for capacitance regulation in the next section.

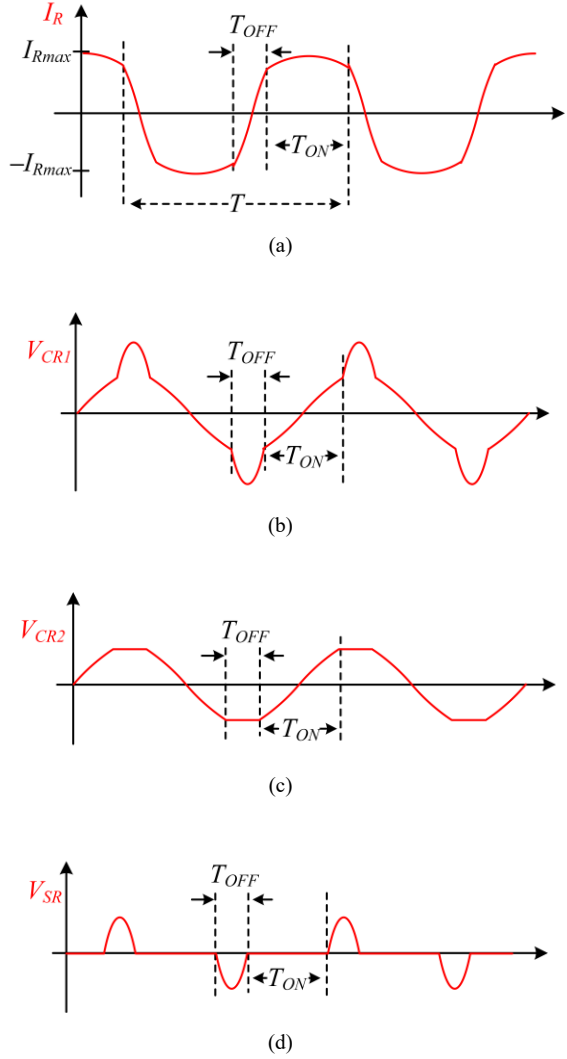

Fig. 3. Waveforms of system voltage and current with the time-division switched-capacitor array. (a) Receiver current. (b) The voltage of the capacitor $C_{R1}$. (c) The voltage of the capacitor $C_{R2}$. (d) The voltage of the switch $S_R$.

### B. Compensation Capacitance Regulation with Time-Division Method

To regulate the equivalent capacitance of the compensation, this paper adopts a time-division regulation method [25], which controls the switch $S_R$ on and off twice during one WPT cycle. As shown in Fig. 3, the capacitor $C_{R1}$ is actively compensating the entire time, while $C_{R2}$ only participates temporarily for $T_{ON}$, twice in one period. Therefore, the effective equivalent capacitance of the compensation $C_{RE}$ can be controlled by switching $S_R$. To be more specific, when the capacitor voltage $V_{CR1}$ increases from 0 to the maximum value, the equivalent capacitance $C_{RE}$ can be expressed as

$$\frac{\int_0^{\frac{T}{4}} I_R dt}{C_{RE}} = \frac{\int_0^{\frac{T_{ON}}{2}} I_R dt}{C_{R1}+C_{R2}} + \frac{\int_{\frac{T_{ON}}{2}}^{\frac{T}{4}} I_R dt}{C_{R1}} \quad (2)$$

where $T$ is the period of one WPT cycle, and $T_{ON}$ is the switch-on time of switch $S_R$ during one control cycle.

As aforementioned, the switch $S_R$ will turn on and off twice. Thus, the control period is half of one WPT period, and $T_{OFF}$ can be expressed as

$$T_{ON}+T_{OFF}=\frac{T}{2}. \quad (3)$$

As the compensation varies, the current $I_R$ in (2) is a piece-wise function and can be expressed as

$$I_R = \begin{cases} I_{Rmax1}\sin\left(2\pi f_T t+\frac{\pi}{2}\right) & \left(t\leq\frac{T_{ON}}{2}\right) \\ I_{Rmax2}\sin\left(2\pi f_T t+\frac{\pi}{2}+\theta_T\right)+K_T & \left(\frac{T_{ON}}{2}\leq t\leq\frac{T}{4}\right) \end{cases} \quad (4)$$

where $\theta_T$ and $K_T$ are offsets unknown yet; $I_{Rmax1}$ and $I_{Rmax2}$ are the maximum currents of the transmitter compensated through paralleled $C_{R1}\|C_{R2}$ and the single capacitor $C_{R1}$, respectively.

For simplification, the current $I_R$ is considered a sinuous waveform and expressed as

$$I_R = I_{Rmax}\sin\left(2\pi f_T t+\frac{\pi}{2}\right) \quad \left(0\leq t\leq\frac{T}{4}\right). \quad (5)$$

Therefore, an approximate analytical solution for $C_{RE}$ can be obtained as

$$C_{RE} = \frac{1}{\dfrac{1-\sin(\pi f_T T_{ON})}{C_{R1}}+\dfrac{\sin(\pi f_T T_{ON})}{C_{R1}+C_{R2}}}. \quad (6)$$

Also, the desired equivalent capacitance can be calculated from [26]

$$C_{RE} = \frac{1}{(2\pi f_T)^2 L_R}. \quad (7)$$

In consequence, with the acknowledged inductance, capacitance, and detected frequency, the switch-on time $T_{ON}$ follows

$$T_{ON} = \frac{1}{\pi f_T}\arcsin\left(\frac{C_{R1}+C_{R2}}{C_{R2}}-(2\pi f_T)^2 L_R\frac{C_{R1}(C_{R1}+C_{R2})}{C_{R2}}\right). \quad (8)$$

After getting the theoretical calculation value of $T_{ON}$, the primary side needs to tune this value because of the difference between (4) and (5). Also, considering the signal $V_{LA}$ may be affected by external electromagnetic interference, the zero-phase point is imprecise. Moreover, if the transmitter current is not constant or load-independent, $I_T$ is generally positively related to the impendence of the receiving circuits $\sum Z_i$, and their relationship can be expressed as [27]

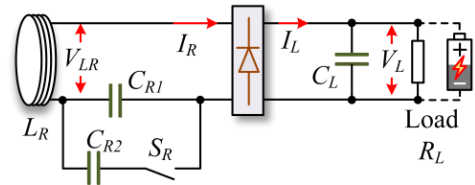

Fig. 4. Receiving circuit with rectifier bridge and load battery.

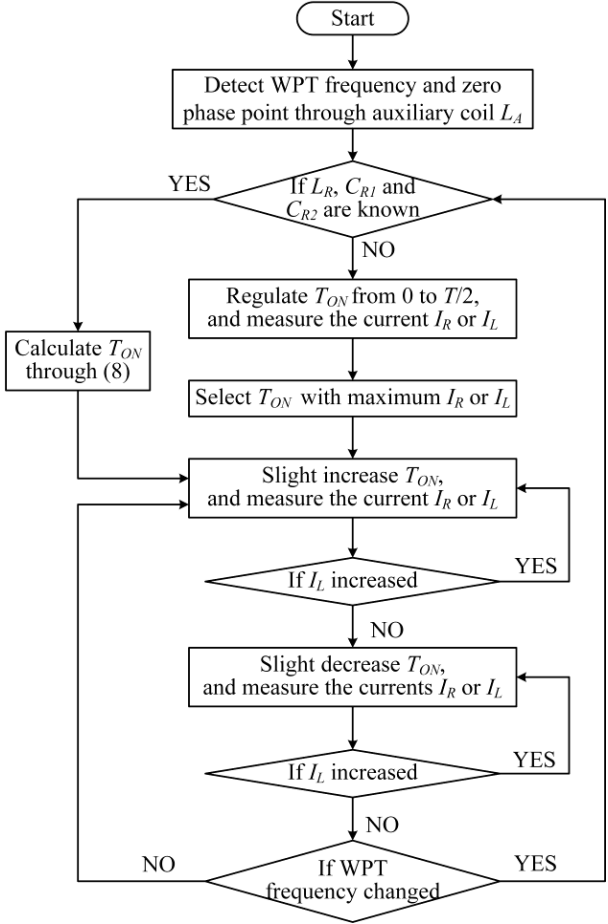

Fig. 5. Flowchart of energy decryption based on time counting and comparison.

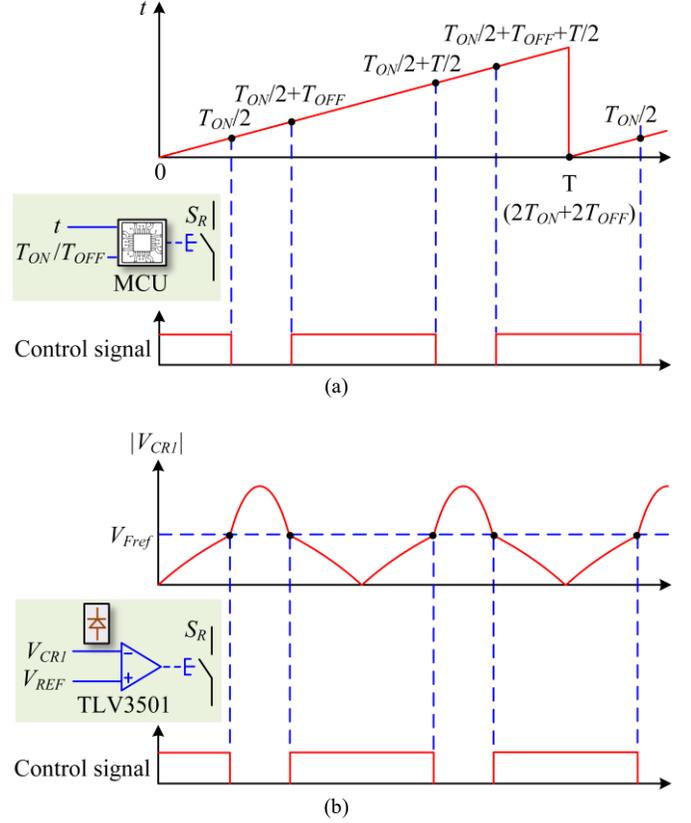

Fig. 6. Control strategy for switch $S_R$. (a) Extract control signal from temporal comparison. (b) Extract control signal from voltage comparison.

$$I_T \propto \frac{V_T}{\sum_{1}^{n} Z'_i} = \frac{V_T}{(2\pi f_T)^2} \sum_{1}^{n} \frac{Z_i}{M_i^2} \quad (9)$$

where $V_T$ is the primary converter's output voltage, and $\sum Z'_i$ is the impedance of all receivers' impedance referred to the transmitter side.

Thus, with larger $I_T$, the $V_{LR}$ may also increase according to (1). In other words, when the current $I_T$ is not load-independent, the maximum power transfer frequency point may drift from the resonant frequency point, and it is very difficult to calculate through equations.

The most applicable way for tuning $T_{ON}/T_{OFF}$ is by detecting the load current or voltage. By increasing or decreasing $T_{ON}$, the load power changes accordingly. Thus, the $T_{ON}$ that relates to the maximum load current or voltage $V_L$ is the desired switching time. It should be mentioned that a large voltage stabilizing capacitor $C_L$ may connect in parallel to the DC-load in some applications, and it will stabilize the voltage and current of the load, as shown in Fig. 4. Also, if the load is a battery, $V_L$ is always increasing during the hacking process. Therefore, it is difficult to detect any $V_L$ variation when changing $T_{ON}$, while the receiver current $I_R$ and the current $I_L$ for both $R_L$ and $C_L$ are the desired reference signals for tuning $T_{ON}$.

To illustrate the proposed energy decryption method better, the flowchart of the whole process is shown in Fig. 5.

## IV. SYSTEM DESIGN

### A. Control Signal Source for Time-Division Switched-Capacitor

There are two possible control strategies for switch $S_R$ to regulate the equivalent capacitance $C_{RE}$: one is based on time counting and comparing with switching time $T_{ON}$, while the other one is based on measuring capacitor voltage $V_{CR1}$ and comparing with voltage threshold $V_{REF}$, as shown in Fig. 6. The one offering higher reliability and possessing better performance will be employed.

For the voltage comparison strategy, the desired voltage threshold $V_{REF}$ can be acquired through negative feedback regulation, as shown in Fig. 7. The largest advantage is that there is no need to detect the WPT frequency or zero-phase point. Also, the voltage comparison can be conducted by a zero-crossing detector (comparator chip). Thus, the controller does not need to measure the high-frequency AC voltage $V_{CR1}$.

However, the feedback control of $V_{REF}$ in Fig. 7 is time-consuming, as $V_{CR1}$ is generally thousands of volts under resonant conditions. Once the reference $V_{REF}$ increases too much and is larger than the maximum value of $V_{CR1}$, $S_R$ will be on all the time; thus, we need to regulate $V_{REF}$ from the beginning. Moreover, if the transmitter current $I_T$, the mutual inductance $M_R$, or the load $R_L$ change, $I_R$ and $V_{CR1}$ follow accordingly. As a result, the intercepting receiver needs to re-regulate the threshold $V_{REF}$.

Therefore, we selected the temporal comparison strategy. Although WPT frequency and zero-phase detection are required, the desired switching time can be acquired faster to deal with variable frequency encryption strategy. Also, if $I_T$ is load-independent, the switching time $T_{ON}/T_{OFF}$ is only related to frequency $f$ according to (8), and there is no need to regulate the switching time when the load changes.

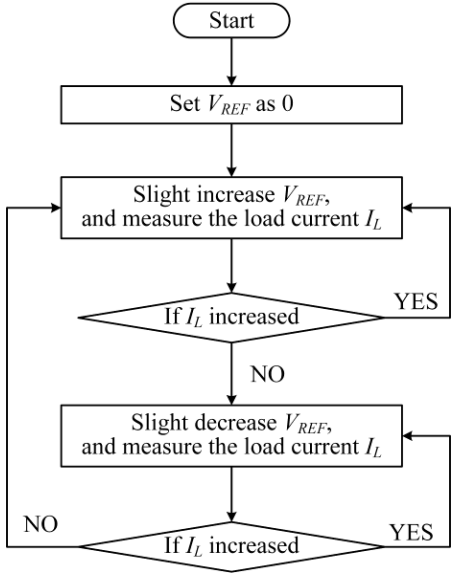

Fig. 7. Flowchart of energy decryption based on voltage measuring and comparison.

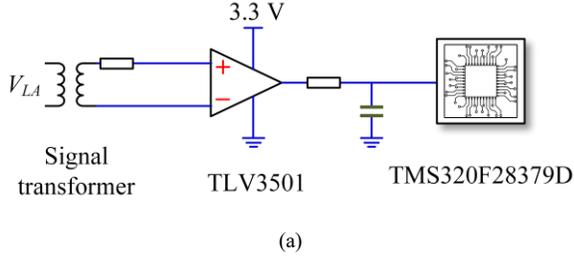

(a)

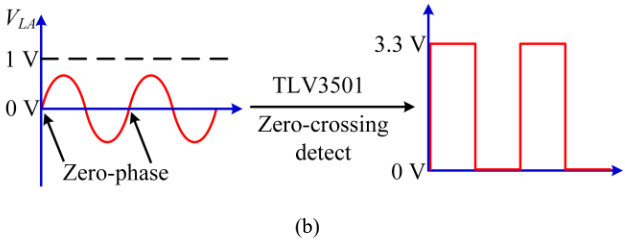

(b)

Fig. 8. Frequency and phase detection. (a) Detection circuit. (b) Signal processing.

### B. Signal Processing Circuit

As the energy we are aiming to hack is generally tens to hundreds of kilohertz, phase detection should be timely and effective.

First, we employed a small signal transformer to isolate the signal $V_{LA}$. Then, considering the signal $V_{LA}$ from coil $L_A$ may be too small to be detected by any signal processor, the TLV3501 thresholds, quantifies, and amplifies $V_{LA}$ with minimum phase lag, as shown in Fig. 8.

### C. Switch for Compensation Capacitor

Previous work with variable capacitors uses only a single MOSFET for switch $S_R$ [28], as shown in Fig. 9(a). However, the capacitor current $I_{R2}$ is AC, while the MOSFET contains a body diode. Thus, the high-frequency capacitor current is only under control during half a period, while the body diode of the MOSFET continuously conducts during the other half-cycle, as shown in Fig. 9(b). Therefore, the adjustment effect is only half of the idea condition, and the fundamental component decreases in this asymmetric voltage waveform.

A pair of reverse-series-connected MOSFETs can solve the problem, as shown in Fig. 9(c). Thus, both the higher and lower half-cycle of the capacitor current can be controlled, as shown in Fig. 9(d).

However, the controller needs to compare the time with four switching times in one cycle, as shown in Fig. 6(a). Especially, if the switch $S_R$ delays turn-on at the second or fourth comparison, the capacitor voltages $V_{CR1}$ and $V_{CR2}$ should be different. Thus, $S_R$ does not turn on at zero voltage, and a large current loop is formed between two capacitors.

Therefore, to avoid the current loop between capacitors and reduce switch-on losses of $S_R$, we employ two MOSFETs and two diodes as $S_R$, as shown in Fig. 9(e) and (f). At the first switching time $T_{ON}/2$, we turn off the first MOSFET, namely $M_1$, and turn on the second MOSFET, namely $M_2$. No current flows through $M_2$ until $V_{CR2}$ is higher than $V_{CR1}$ because of $D_2$. Hence, $S_R$ will be on automatically and preciously at the second switching time $T_{ON}/2+T_{OFF}$. It should be mentioned that both $D_2$ and $M_2$ switch on at zero voltage, just like shown in Fig. 3(d), so they are well protected. Similarly, at the third switching time $T_{ON}/2+T/2$, $M_2$ is off and $M_1$ is on. Then, $S_R$ will be on automatically and preciously at the fourth switching time, and both $D_1$ and $M_1$ switch on at zero voltage.

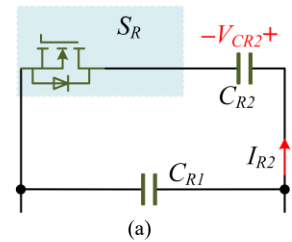

(a)

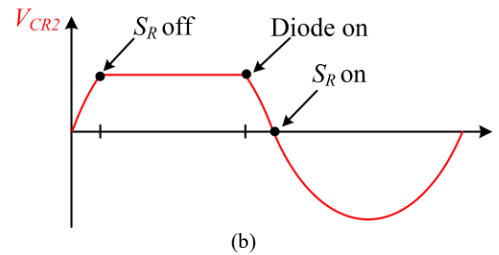

(b)

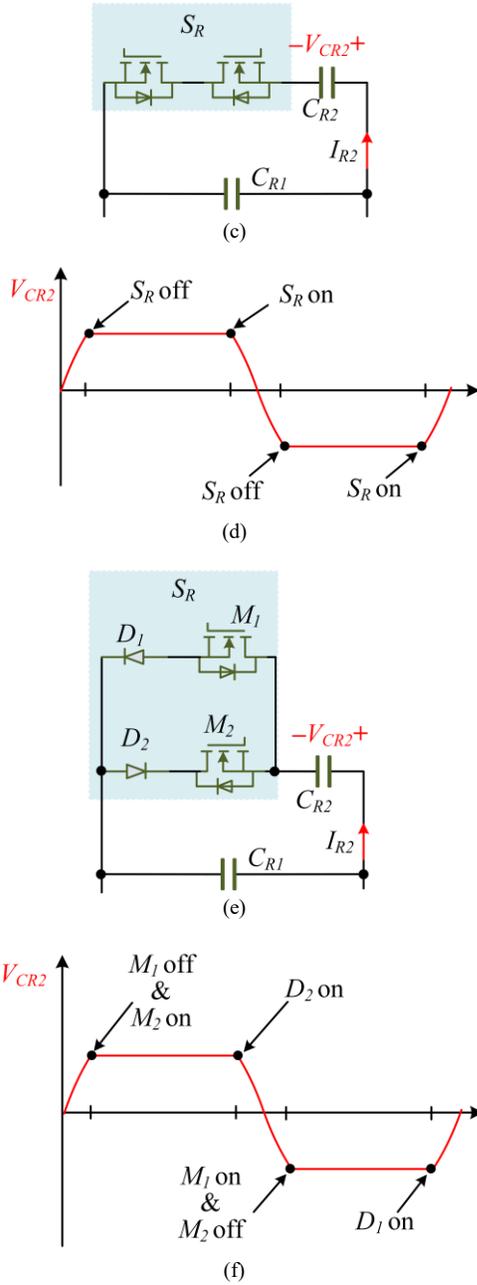

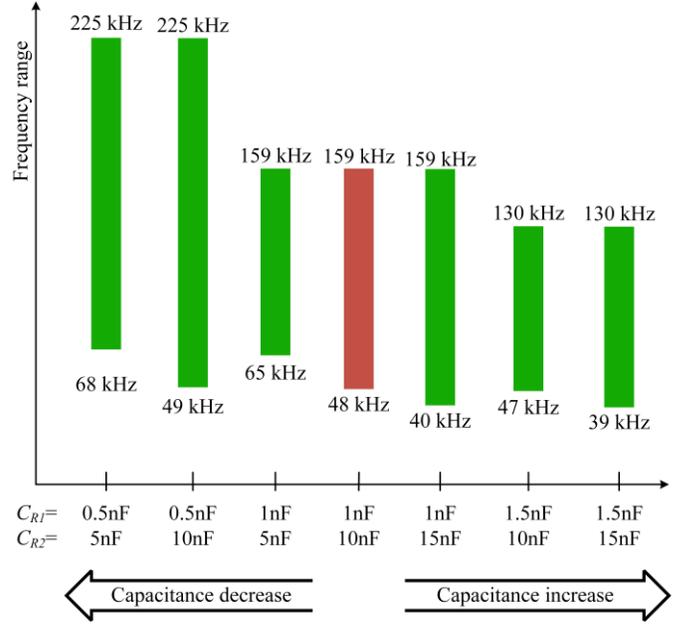

Fig. 9. Various circuits for adaptive capacitor compensation. (a) Single transistor for the capacitor. (b) Asymmetric capacitor voltage. (c) Double transistor as the bidirectional switch for the capacitor. (d) Symmetrical capacitor voltage with four control points. (e) Parallel-transistor-and-diode circuit for switch $S_R$. (f) Symmetrical capacitor voltage with two control points.

### D. Capacitance Selection

The switched-capacitor array allows changing the equivalent capacitance $C_{RE}$ in the range of $[C_{R1}, C_{R1}+C_{R2}]$. Thus, the hacking frequency range reaches

$$\frac{1}{2\pi\sqrt{L_R(C_{R1}+C_{R2})}} \leq f_R \leq \frac{1}{2\pi\sqrt{L_R C_{R1}}}. \tag{10}$$

Therefore, $C_{R1}$ should be low enough to compensate the receiver at the highest frequency $f_H$, while $C_{R1}+C_{R2}$ should be large enough to compensate $L_R$ at the lowest frequency $f_L$.

Fig. 10. Relationship between the hacking frequency range and capacitance variation.

As a result, capacitances $C_{R1}$ and $C_{R2}$ can be expressed as

$$\begin{cases} C_{R1} \leq \dfrac{1}{(2\pi \times f_H)^2 L_R} \\ C_{R2} \geq \dfrac{1}{(2\pi \times f_L)^2 L_R} - C_{R1} \end{cases}. \tag{11}$$

Generally, the international standard of automotive wireless electricity transmission of SAE is 85 kHz [29]. Thus, the decryption frequency range should include at least 80~100 kHz. Then, the capacitances can be selected through (11).

As various capacitor types have aging problems and capacitances tend to decrease, it is desired to select a larger capacitance of $C_{R2}$. Otherwise, the lower limit of the hacking frequency may increase over time. To be more specific, if the inductance $L_R$ is constant as 1 mH, while capacitances $C_{R1}$ and $C_{R2}$ change from 50% to 150%, the hacking frequency range will change as shown in Fig. 10. Also, this figure illustrates that if $C_{R1}$ is too large, the upper limit $f_H$ may not be high enough. Thus, it is desired to select a smaller capacitance of $C_{R1}$ to avoid an insufficient upper limit $f_H$.

Meanwhile, $C_{R2}$ cannot be infinitely large, otherwise, the equivalent capacitance $C_{RE}$ will be too sensitive to $T_{ON}/T_{OFF}$. Also, $C_{R1}$ cannot be infinitesimal. Otherwise, the peak voltage of $V_{CR1}$ would be too high. In consequence, the peak voltage may exceed the limits of switch $S_R$ and break it.

TABLE I
SIMULATION PARAMETERS

| Item | Value/Type | Unit |
|---|---|---|
| WPT frequency range ($f_T$) | 50~300 | kHz |
| Transmitter coil inductances ($L_T$) | 150 | μH |
| Receiver coil inductances ($L_R$, $L_{R50}$, $L_{R300}$) | 80, 80, 80 | μH |
| Compensation capacitances ($C_{R1}$, $C_{R2}$) | 3, 130 | nF |
| Load resistance ($R_L$) | 25 | Ω |

## V. SIMULATION

For verification, a series of computational simulations in MATLAB/Simulink provided verification. Transmitter $L_T$ offers load-independent current $I_T$ from 50~300 kHz. In addition to the transmitter and hacking receiver, two more receivers $L_{R50}$ and $L_{R300}$ serve for comparison. The resonant frequencies for low-frequency receiver $L_{R50}$ and high-frequency receiver $L_{R300}$ were 50 kHz and 300 kHz, respectively, and other key parameters are given in Table I.

When the transmitter provides $I_T$ at 50 kHz, 120 kHz, and 300 kHz, the hacking controller calculates $T_{ON}$ as 9.69 μs, 2.87 μs, and 0.16 μs respectively, as shown in Fig. 11.

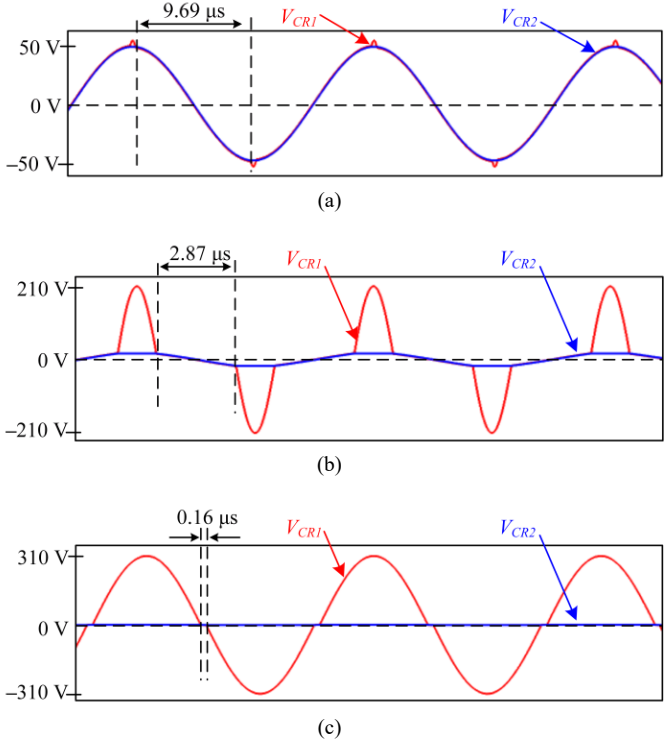

Fig. 11. Voltages of the capacitors for the hacking receiver (a) at 50 kHz, (b) at 120 kHz, and (c) at 300 kHz.

For a comparison, Fig. 12 shows currents of $L_R$, $L_{R50}$, and $L_{R300}$, namely $I_R$, $I_{R50}$, and $I_{R300}$, at different frequencies. The frequency detection and feedback control processes are omitted, and $T_{ON}$ is directly acquired from (8) to put three-frequency stages into one figure. It proves that, even only employing coarse tuning, the current of the hacking receiver current is only slightly smaller than the full-resonant receiver current and much larger than the un-resonant receiver current. Moreover, the ratio of the loads' power, namely $P_R$: $P_{R50}$: $P_{R300}$, is 0.97: 1: 0.001, 1: 0.28: 0.01, and 0.88: 0.03: 1, at 50 kHz, 120 kHz, and 300 kHz, respectively.

Therefore, the hacking receiver can steal substantial amounts of energy from the transmitter for a wide frequency range with the calculated $T_{ON}$ from (8).

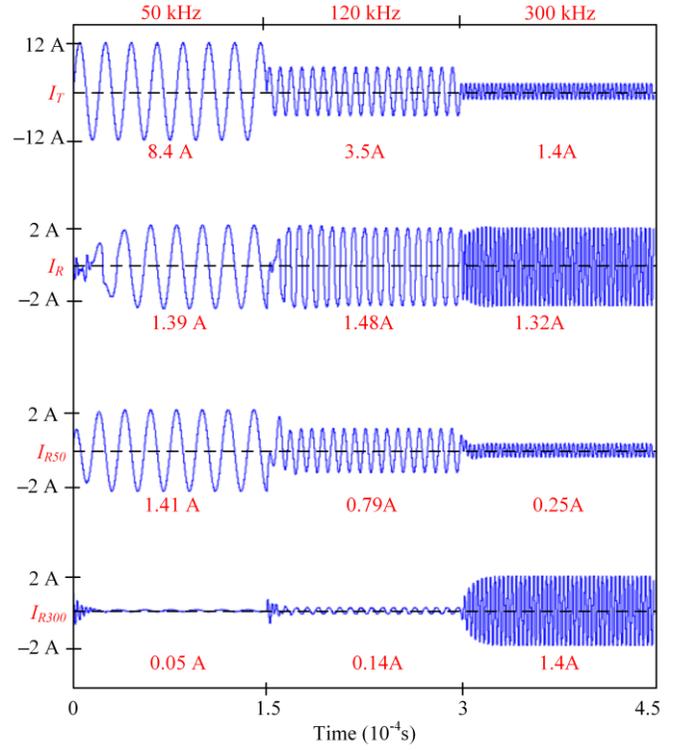

Fig. 12. Currents of the transceiver and the three receivers at different frequencies: from top to bottom (1) the transmitter, (2) the unauthorized interceptor, (3) the receiver with fixed resonance at 50 kHz, and (4) the receiver with fixed resonance at 300 kHz.

TABLE II
EXPERIMENT PARAMETERS

| Item | Value/Type | Unit |
|---|---|---|
| WPT frequency range ($f_T$) | 79~161 | kHz |
| Transmitter coil inductances ($L_T$) | 150 | μH |
| Receiver coil inductances ($L_R$, $L_{R79}$, $L_{R161}$) | 80, 80, 80 | μH |
| Auxiliary coil inductances ($L_A$) | 10 | μH |
| Mutual inductances among transmitter and receivers and auxiliary coil ($M_R$, $M_A$, $M_{RA}$) | 15, 3, 2 | μH |
| Compensation capacitances ($C_{R1}$, $C_{R2}$) | 10, 44 | nF |
| Load resistances ($R_L$, $R_{79}$, $R_{161}$) | 25, 25, 25 | Ω |
| Transmission distance | 40 | mm |
| Diameter of the transmitter coil | 19 | mm |
| Diameter of the receiver coil | 9 | mm |

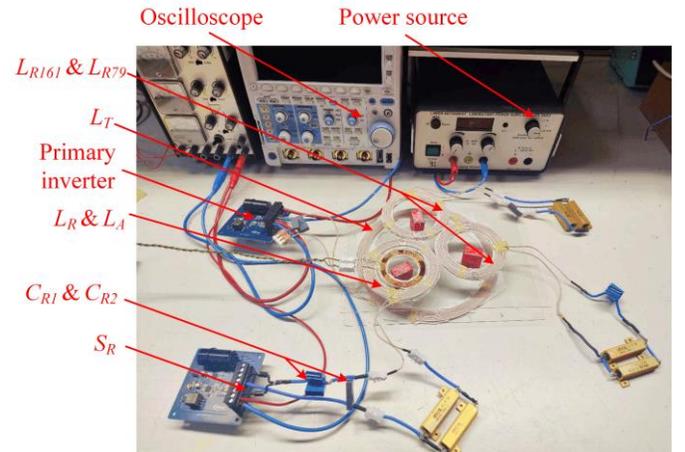

Fig. 13. Experimental setup.

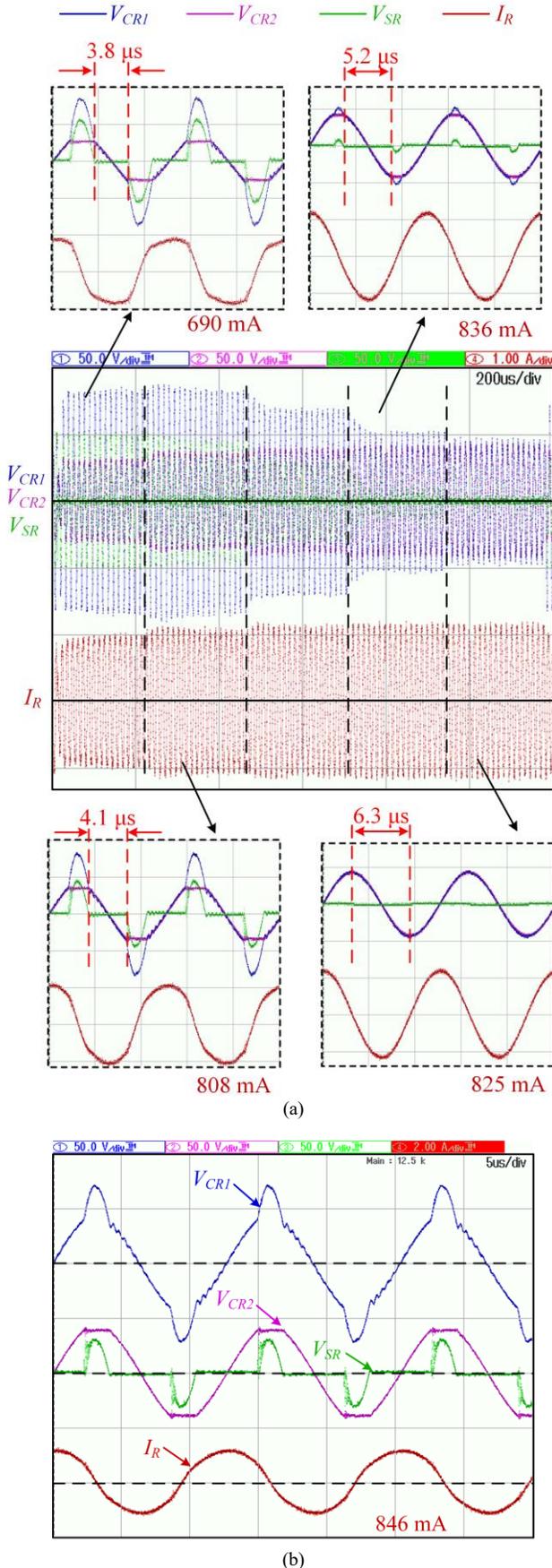

Fig. 14. Voltages and current of the hacking receiver under 79 kHz. (a) Tuning $T_{ON}$. (b) With the desired $T_{ON}$.

## VI. EXPERIMENT

We performed experiments to verify the effectiveness of the proposed energy decryption strategy, as shown in Fig. 13. The experimental setup contains one transmitter, three receivers, and one auxiliary coil, and key parameters are given in Table II. The hacking receiver $L_R$ can harvest energy from 79 kHz to 161 kHz, while the other two receivers $L_{R79}$ and $L_{R161}$ are resonant at 79 kHz and 161 kHz, respectively.

When the transmitter provides 3.4 A current at 79 kHz, the auxiliary coil detects and transfers the frequency and zero-phase points to the controller. Then, the hacking controller calculates $T_{ON}$ as 4.3 μs according to (5) and (8), and controls the shutdown time of the switch $S_R$ to de-active $C_{R2}$, as shown in Fig. 14.

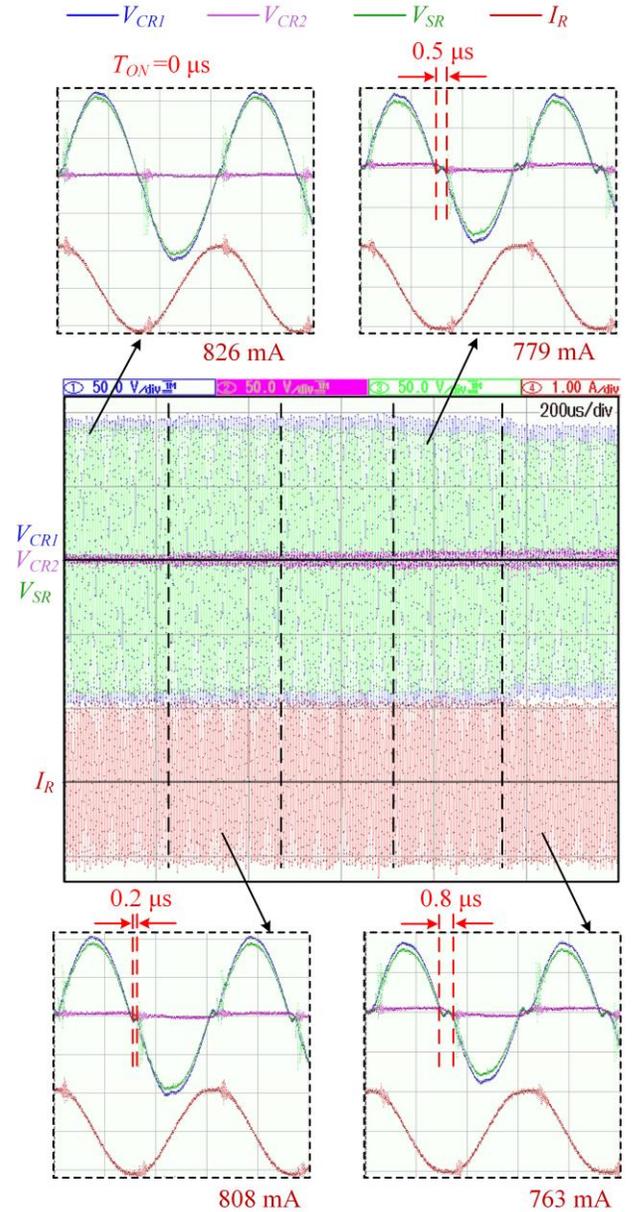

Fig. 15. Voltages and current of the hacking receiver under 161 kHz.

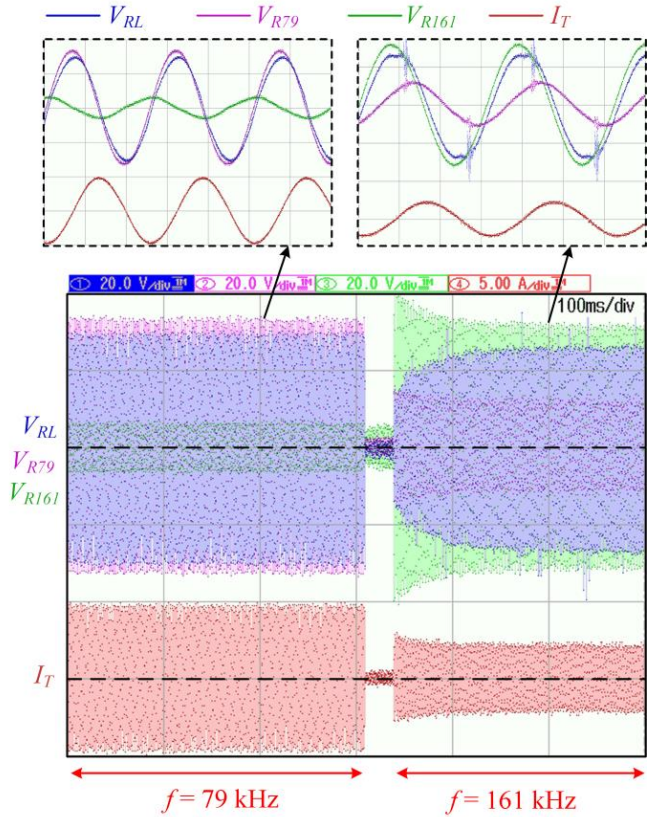

Fig. 16. Load voltages under different frequencies.

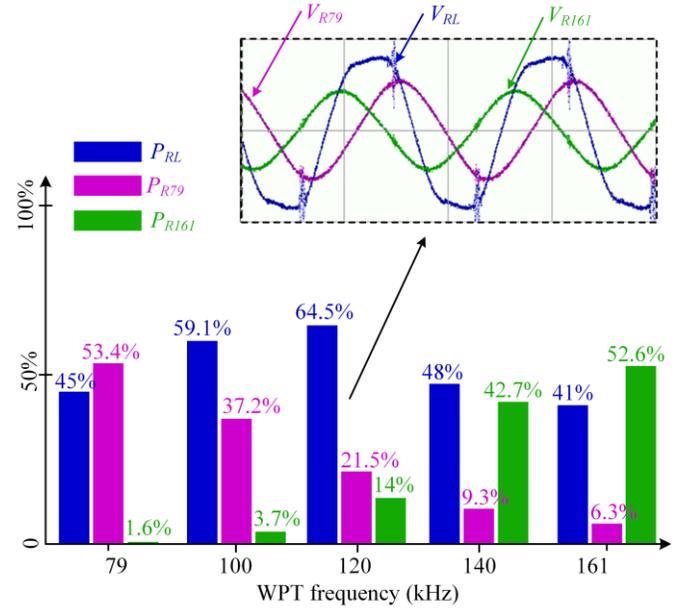

Fig. 17. Percent of energy received by each receiver under different frequencies.

To avoid the potential error, the hacking receiver needs to tune $T_{ON}$ around the calculated result and detect the corresponding $I_R$. When increasing $T_{ON}$ from 3.8 to 6.3 μs, the rated value of $I_R$ increases from 690 to 846 mA, and then decreases to 830 mA, as shown in Fig. 14 (a). Thus, 4.7 μs is finally adopted as the best $T_{ON}$ for the current magnetic field, and the corresponding voltages of compensation capacitors and switch are shown in Fig. 14 (b).

However, when the transmitter provides 1.7 A $I_T$ at 161 kHz, $C_{R2}$ only participates in the compensation for a short time to achieve a low $C_{RE}$, as shown in Fig. 15. Since the calculated result is 0.22 μs according to (8), the controller detects $I_R$ when regulating $T_{ON}$ from 0 to 0.8 μs. The results show 0 μs is the optimized result for $T_{ON}$ under 160 kHz. It should be mentioned that all MOSFETs and diodes suffer from leakage current. Thus, even when $S_R$ is off all the time, $C_{R2}$ still participates in the compensation; the real compensation $C_{RE}$ would therefore be slightly higher than $C_{R1}$.

For comparison, the load voltage $V_{RL}$ and load power $P_{RL}$ are compared with peers of low and high-frequency receivers, namely, $V_{R79}$, $V_{R161}$, $P_{R79}$, and $P_{R161}$, respectively. As shown in Fig. 16, when the transmitter frequency is 79 kHz, the rated values of $V_{RL}$, $V_{R79}$, and $V_{R161}$ are 21.7 V, 23.7 V, and 4.1 V, respectively. Thus, the ratio of receiving power among the hacking receiver, low-frequency receiver, and high-frequency receiver should be 0.84: 1: 0.03. However, when the transmitter changes the frequency $f$ to 161 kHz. The rated values of $V_{RL}$, $V_{R79}$, and $V_{R161}$ become 19.7 V, 7.6 V, and 22.3 V, respectively, and the power ratio $P_{RL}$: $P_{R79}$: $P_{R161}$ becomes 0.78: 0.12: 1.

Moreover, Fig. 17 illustrates the proportion of energy received by each receiver to the total received energy from 79 kHz to 161 kHz. It proves that the proposed energy decryption method can steal substantially more energy than non-resonant receivers.

The identification of the new frequency requires several milliseconds, and then the unauthorized interceptor finishes the whole tuning process within 100 ms, which is only slightly slower than a fixed compensation to build up the oscillation after a frequency change.

## VII. CONCLUSION

This paper proposes an energy decryption method for frequency-varying encrypted WPT systems. Prominently, only two capacitors and one high-frequency switch are required. The key to adaptive resonance and impedance is controlling the duty cycle of $C_{R2}$ in one cycle in the compensation to achieve a desired equivalent capacitance for different and variable WPT frequencies. Both simulation and experiments prove that the proposed method works well from 79 kHz to 161 kHz, and the hacking receiver can successfully harvest about 78–84% energy of the full-resonant receiver under the same condition. It should be mentioned that even eventually hacking very little energy is already a serious problem. The presented hacking system, however, can even continuously and massively steal energy. Significantly, this energy decryption method can provide a general way to design an energy harvester and energy capturer from electromagnetic fields of WPT systems.

Future research may study high-order LC and LCC compensation to protect the transceiver from sudden over-load through load-independent load current [30, 31]. Dynamically changing LCC compensation would require control methods for two switched-capacitor arrays. Furthermore, strong harmonics and interference may affect the hacking transceiver if they are on a similar amplitude order of magnitude as the main signal, which future attacks may consider to increase the stability of energy theft in such cases.

AUTHOR CONTRIBUTIONS

SMG designed the research as well as the concept of wireless power encryption through and secured funding. HW performed the work and conceived the tunable compensation. NT, WJ, CQJ consulted on electronics, controller specifics, and wireless power. SMG and HW wrote the text.